\documentclass{llncs}

\usepackage{subfig}
\usepackage[latin1]{inputenc}
\usepackage{graphicx}
\usepackage{amsmath}
\usepackage{url}

\usepackage{algorithmic}
\usepackage{algorithm}

\begin{document}

\title{A Bit-Compatible Shared Memory Parallelization for ILU(k) Preconditioning and a Bit-Compatible Generalization to Distributed Memory}
\author{Xin Dong\thanks{This work was partially supported by
               the National Science Foundation under Grants
                CCF 09-16133 and CNS-06-19616.} 
 \and Gene Cooperman$^\star$}
\institute{College of Computer Science, Northeastern University\\
 Boston, MA 02115, USA\\
\email{\{xindong,gene\}@ccs.neu.edu}\\
}
\maketitle
\begin{abstract}
  ILU(k) is a commonly used preconditioner for iterative linear
  solvers for sparse, non-symmetric systems.  It is often preferred
  for the sake of its stability.  We present TPILU(k), the first
  efficiently parallelized ILU(k) preconditioner that maintains this
  important stability property.  Even better, TPILU(k) preconditioning
  produces an answer that is bit-compatible with the sequential ILU(k)
  preconditioning.  In terms of performance, the TPILU(k)
  preconditioning is shown to run faster whenever more cores are made
  available to it --- while continuing to be as stable as sequential
  ILU(k).  This is in contrast to some competing methods that may
  become unstable if the degree of thread parallelism is raised too
  far.  Where Block Jacobi ILU(k) fails in an application, it can be
  replaced by TPILU(k) in order to maintain good performance, while
  also achieving full stability.  As a further optimization, TPILU(k)
  offers an optional {\em level-based incomplete inverse method} as a
  fast approximation for the original ILU(k) preconditioned matrix.
  Although this enhancement is not bit-compatible with classical
  ILU(k), it is bit-compatible with the output from the
  single-threaded version of the same algorithm.  In experiments on a
  16-core computer, the enhanced TPILU(k)-based iterative linear
  solver performed up to 9~times faster.  As we approach an era of
  many-core computing, the ability to efficiently take advantage of
  many cores will become ever more important. TPILU(k) also
  demonstrates good performance on cluster or Grid. For example, the
  new algorithm achieves 50 times speedup with 80 nodes for general
  sparse matrices of dimension 160,000 that are diagonally dominant.
\end{abstract}

\section{Introduction}
\label{Introduction}
This work introduces a parallel preconditioner, TPILU(k), with good
stability and performance across a range of sparse, non-symmetric
linear systems.  For a large sparse linear system~$Ax=b$, parallel
iterative solvers based on
ILU(k)~\cite{springerlink:10.1007/BF01931691,WATTS} often suffer from
instability or performance degradation.  In particular, most of
today's commonly used algorithms are domain decomposition
preconditioners, which become slow or unstable with greater
parallelism.  This happens as they attempt to approximate a linear
system by more and smaller subdomains to provide the parallel work for
an increasing number of threads.  The restriction to subdomains of
ever smaller dimension must either ignore more of the off-diagonal
matrix elements, or must raise the complexity by including
off-diagonals into the computation for an optimal decomposition.  The
former tends to create instability for large numbers of threads (i.e.,
for small subdomains), and the latter is slow.

Consider the parallel preconditioner
PILU~\cite{Hysom00ascalable,HysomPothen03} as an example. PILU would
experience performance degradation unless the matrix $A$ is {\em
  well-partitionable} into subdomains. This condition is violated by
linear systems generating many fill-ins (as occurs with higher initial
density or higher level $k$) or by linear solvers employing many
threads. Another parallel preconditioner BJILU~\cite{SAADBOOK} (Block
Jacobi ILU(k)), would fail to converge as the number of threads~$w$
grows. This is especially true for linear systems that are not
diagonally dominant, in which the solver might become invalid by
ignoring significant off-diagonal entries. This kind of performance
degradation or instability is inconsistent with the widespread
acceptance of parallel ILU(k) for varying~$k$ to provide efficient
preconditioners.

In contrast, TPILU(k) is as stable as sequential ILU(k) and its
performance increases with the number of cores.  TPILU(k) can capture
both properties simultaneously --- precisely because it is not based
on domain decomposition.  In the rest of this paper, we will simply
write that {\em TPILU(k) is stable} as a shortened version of the
statement that TPILU(k) is stable for any number of threads whenever
sequential ILU(k) is stable.

TPILU(k) uses a task-oriented parallel ILU(k) preconditioner for the
base algorithm.  However, it optionally first tries a different,
level-based incomplete inverse submethod~({\em TPIILU(k)}).  The term
{\em level-based incomplete inverse} is used to distinguish it from
previous methods such as ``threshold-based'' incomplete
inverses~\cite{Bollhofer:2002:RIF:587705.587781}.  The level-based
submethod either succeeds or else it fails to converge.  If it doesn't
converge fast, TPILU(k) quickly reverts to the stable, base
task-oriented parallel ILU(k) algorithm.

A central point of novelty of this work concerns
bit-compatibility.  The base task-oriented parallel component of
TPILU(k) is bit-compatible with classical sequential ILU(k), and the
level-based optimization produces a new algorithm that is also
bit-compatible with the single-threaded version of that same
algorithm.  Few numerical parallel implementations can guarantee
this stringent standard.  The order of operations is precisely
maintained so that the low order bits due to round-off do not change
under parallelization.  Further, the output remains bit-compatible as
the number of threads increases --- thus eliminating worries whether
scaling a computation will bring increased round-off error.

In practice, bit-compatible algorithms are well-received in the
workplace.  A new bit-compatible version of code may be substituted
with little discussion.  In contrast, new versions of code that result
in output with modified low-order bits must be validated by a
numerical analyst.  New versions of code that claim to produce more
accurate output must be validated by a domain expert.

A prerequisite for an efficient implementation in this work was the
use of thread-private memory allocation arenas.  The implementation
derives from~\cite{xingenejohn}, where we first noted the issue.  The
essence of the issue is that any implementation of POSIX-standard
``malloc'' libraries must be prepared for the case that a second
thread frees memory originally allocated by a first thread.  This
requires a centralized data structure, which is slow in many-core
architectures.  Where it is known that memory allocated by a thread
will be freed by that same thread, one can use a thread-private
(per-thread) memory allocation arena.  The issue arises in the memory
allocations for ``fill-ins'' for symbolic factorization.  In
LU-factorization based algorithms, the issue is still more serious
than incomplete~LU, since symbolic factorization is a relatively
larger part of the overall algorithm.

TPILU(k) is generalized to computer clusters, which supports a hybrid
memory model using multiple computers, each with multiple cores. This
helps to further improve the performance for the TPILU(k) algorithm by
aggregating cores from many computers when the computation is highly
intensive. A pipeline communication model is employed for efficient
local network usage, which overlaps communication and computation
successfully given the number of computers is not huge.

The rest of this paper is organized as follows.
Section~\ref{sec:ReviewILUK} reviews LU factorization and sequential
ILU(k) algorithm. Section~\ref{sec:TOPILUK} presents task-oriented
parallel TPILU(k), including the base algorithm
(Sections~\ref{sec:BeginBaseAlgo} through~\ref{sec:EndBaseAlgo}) and
the level-based incomplete inverse submethod
(Section~\ref{sec:IncompleteInverse}).  Section~\ref{sec:Experiment}
analyzes the experimental results.  We review related work in
Section~\ref{RELATEDWORK}.

\section{Review of the Sequential ILU(k) Algorithm}
\label{sec:ReviewILUK}
A brief sketch is provided.  See~\cite{SAAD2} for a detailed review of
ILU(k).  LU~factorization decomposes a matrix~$A$ into the product of
a lower triangular matrix~$L$ and an upper triangular matrix~$U$.
From~$L$ and~$U$, one efficiently computes $A^{-1}$ as $U^{-1}L^{-1}$.
While computation of~$L$ and~$U$ requires $O(n^3)$ steps, once
done, the computation of the inverse of the triangular matrices
proceeds in $O(n^2)$ steps.

For sparse matrices, one contents oneself with solving~$x$ in $Ax=b$
for vectors $x$ and~$b$, since $A^{-1}$, $L$ and~$U$ would all be
hopelessly dense.  Iterative solvers are often used for this purpose.
An ILU(k) algorithm finds sparse approximations, $\widetilde L\approx
L$ and $\widetilde U\approx U$.  The preconditioned iterative solver
then implicitly solves $A \widetilde U^{-1} \widetilde L^{-1}$, which
is close to the identity.  For this purpose, triangular solve
operations are integrated into each iteration to obtain a solution $y$
such that
\begin{equation}
\label{eq:triangleSolve}
\begin{split}
\widetilde L \widetilde U y = p \\
\end{split}
\end{equation}
where $p$ varies for each
iteration.  This has faster convergence and better numerical
stability.  Here, the {\em level limit} $k$ controls how many
elements should be computed in the process of incomplete LU
factorization.  A level limit of~$k=\infty$ yields full LU-factorization.

Similarly to LU~factorization, ILU(k) factorization can be
implemented by the same procedure as Gaussian elimination. Moreover,
it also records the elements of a lower triangular matrix
$\widetilde L$. Because the diagonal elements of $\widetilde L$ are defined
to be~1, we do not need to store them. Therefore, a single {\em filled
  matrix}~$F$ is sufficient to store both $\widetilde L$
and~$\widetilde U$.

\subsection {Terminology for ILU(k)}
\label{sec:Definition}
For a huge sparse matrix, a standard dense format would be wasteful.
Instead, we just store the position and the value of non-zero
elements.  Similarly, incomplete LU factorization does not insert all
elements that are generated in the process of factorization.  Instead,
it employs some mechanisms to control how many elements are
stored. ILU(k)~\cite{springerlink:10.1007/BF01931691,WATTS} uses the
level limit $k$ as the parameter to implement a more flexible
mechanism.  We next review some definitions.

{\bf \noindent Definition 2.1:} {\em
{\em A fill entry}, or {\em entry} for short, is an element
  stored in memory. (Elements that are not stored are called zero
  elements.)}

{\bf \noindent Definition 2.2:} {\em
{\em Fill-in}:
  Consider Figure~\ref{fig:fill-in}. If there exists $h$ such that
  $i, j>h$ and both $f_{ih}$ and $f_{hj}$ are fill entries, then the
  ILU(k) factorization algorithm may fill in a non-zero value when
  considering rows $i$ and~$j$.  Hence, this
  element~$f_{ij}$ is called a {\em fill-in}; i.e., an entry candidate.
  We say the fill-in~$f_{ij}$ is {\em caused} by the existence
  of the two entries~$f_{ih}$ and~$f_{hj}$.  The entries $f_{ih}$
  and~$f_{hj}$ are the {\em causative entries} of~$f_{ij}$.
  The causality will be made clearer in the next subsection.}

{\bf \noindent Definition 2.3:} {\em {\em Level}: Each entry~$f_{ij}$
  is associated with a level, denoted as $level~(i, j)$ and defined
  recursively by
\begin{equation*}
level~(i,j)=
\begin{cases}
0, & \text{if } a_{ij} \ne 0\\
\min_{1 \le h < \min{(i,j})} level~(i,h)+level~(h,j)+1, & \text{otherwise}
\end{cases}
\end{equation*}
}

The {\em level limit}~$k$ is used to control how many fill-ins
should be inserted into the filled matrix during ILU(k) factorization.
Those fill-ins with a level smaller than or equal to~$k$ are
inserted into the filled matrix~$F$.  Other fill-ins are ignored.
By limiting fill-ins to level~$k$ or less, ILU(k) maintains a sparse
filled matrix.

\subsection {ILU(k) Algorithm and its Parallelization}
\label{sec:Algorithm}
For LU factorization, the defining equation $A=LU$ is expanded into
$a_{ij}=\sum_{h=1}^{min(i,j)}l_{ih}u_{hj}$, since $l_{ih}=0$ for $i>j$
and $u_{hj}=0$ for $i<j$.  When $i > j$, $f_{ij}=l_{ij}$ and we can
write $a_{ij}=\sum_{h=1}^{j-1}l_{ih}u_{hj} + f_{ij}u_{jj}$.  When
$i\le j$, $f_{ij}=u_{ij}$ and we can write
$a_{ij}=\left(\sum_{h=1}^{i-1}l_{ih}u_{hj}\right) +
l_{ii}f_{ij}=\left(\sum_{h=1}^{i-1}l_{ih}u_{hj}\right) + f_{ij}$.
Rewriting them yields the equations for LU~factorization.
\begin{equation}
\label{LUFACTORIZATION}
f_{ij} = 
\begin{cases}
\left(a_{ij} - \sum_{h=1}^{j-1}l_{ih} u_{hj}\right)/u_{jj}, & i>j\\
a_{ij} - \sum_{h=1}^{i-1}l_{ih} u_{hj}, & i\le j 
\end{cases}
\end{equation}

\begin{figure}
\centering
\subfloat[Causative Relationship]{\label{fig:fill-in}\includegraphics[scale=1]{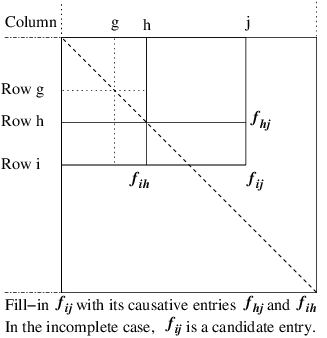}}
~~~~
\subfloat[View of Matrix as Bands]{\label{fig:Viewbands}\includegraphics[scale=1]{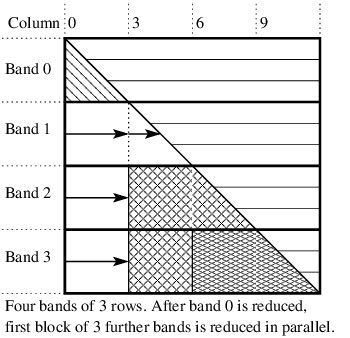}}
\caption{Parallel Incomplete LU Factorization}
\label{ILUFACTORIZATION}
\end{figure}

The computation for incomplete LU factorization follows a definition
similar to Equations~\ref{LUFACTORIZATION} except it skips zero
elements. In our implementation, the matrix~$F$ is initialized to~$A$
and stored in row-major order form prior to any computation. The
computation can be re-organized to use the above equations in the
forward direction.  As each term $l_{ih} u_{hj}$ for $h<j$ is
determined, it can immediately be subtracted from $f_{ij}$. Just as
all of the remaining rows can be reduced simultaneously by the first
row in Gaussian elimination, a row-major order for ILU(k)
factorization leads to a natural parallel algorithm.

Following the defining equations, the ILU(k) algorithm maintains in
memory two rows: row~$h$ and row~$i$, where $h < i$.  Row~$h$ is used
to {\em partially reduce} row~$i$.  For each possible~$j$, the product
$l_{ih}u_{hj}$ is used to reduce the entry $f_{ij}$. Once we have
accumulated all products $l_{ih}u_{hj}$ for $h < min(i,j)$, we are
done.

ILU(k) is separated into two passes: {\em symbolic factorization} or
Phase~I to compute the levels and insert all fill-ins with the level
less than or equal to the level limit $k$ into the filled matrix; and
{\em numeric factorization} or Phase~II to compute the values of all
entries in the filled matrix. Both passes follow a procedure similar
to that described above.  Algorithm~\ref{alg:Levelsub} illustrates the
symbolic factorization phase. It determines for each row~$j$, the set
of permitted entries, $permitted(j)$.  These are the entries for which
the computed entry level or {\em weight} is less than or equal to
the~$k$.  Numeric factorization is simpler, but similar in spirit to
the row-merge update pass of Algorithm~\ref{alg:Levelsub}.  The
lines~14 through~17 control the entries to be updated, and the update
in line~19 is replaced by an update numeric value.  The details are
omitted.

\begin{algorithm}
\begin{algorithmic}[1]
\STATE //Calculate levels and permitted entry positions
\STATE {//Loop over rows}
\FOR{ $j = 1$ to $n$}
  \STATE //Initialization: admit entries in A, and assign them the level zero.
  \STATE $permitted(j)\leftarrow$ empty set  //permitted entry in row $j$
  \FOR{ $t$ = 1 to $n$  // nonzero entries in row~$j$}
    \IF{ $A_{j,t} \not= 0$}
      \STATE $level(j,t)\leftarrow 0$
      \STATE insert $t$ into $permitted(j)$
\ENDIF
\ENDFOR
\ENDFOR
\STATE //Row-merge update pass
\FOR{ each unprocessed $i\in permitted(j)$ with $i
             < j$, in ascending order}
  \FOR{ $t\in permitted(i)$ with $t > i$}
    \STATE $weight =$ $level(j,i)$ + $level(i,t)$ + 1
      \IF{ if $t \in permitted(j)$}
        \STATE  //already nonzero in $F_{j,t}$
        \STATE $level(j,t) \leftarrow \min\{level(j,t), weight\}$
       \ELSE
         \STATE //zero in $F_{j,t}$
         \IF{ $weight \le k$ \qquad //level control}
           \STATE insert $t$ into $permitted(j)$
           \STATE $level(j,t) \leftarrow weight$
          \ENDIF
        \ENDIF
  \ENDFOR
\ENDFOR
\STATE \textbf{return} $permitted$
\end{algorithmic}
\caption{Symbolic factorization:  Phase~I of ILU(k) preconditioning}
\label{alg:Levelsub}
\end{algorithm}

The algorithm has some of the same spirit as Gaussian elimination
if one thinks of ILU(k) as using the earlier row~$h$ to {\em reduce} the
later row~$i$.  This is the crucial insight in the parallel ILU(k)
algorithm of this paper.  One splits the rows of~$F$ into bands,
and reduces the rows of a later band by the rows of an earlier band.
Distinct threads can reduce distinct bands simultaneously, as illustrated
in Figure~\ref{fig:Viewbands}.

\section{TPILU(k):  Task-oriented Parallel ILU(k) Algorithm}
\label{sec:TOPILUK}
\subsection{Parallel Tasks and Static Load Balancing}
\label{sec:BeginBaseAlgo}
We introduce the following definition to describe a general parallel
model, which is valid for Gaussian elimination as well as ILU(k) and
ILUT~\cite{SAADBOOK}.

{\bf \noindent Definition 3.1:} The {\em frontier} is the maximum
number of rows that are currently factored completely.

According to this definition, the frontier $i$ is the limit up to
which the remaining rows can be partially factored except for the
$(i+1)^{th}$ row. The $(i+1)^{th}$ row can be factored completely.
That changes the frontier to $i+1$.

Threads synchronize on the frontier. To balance and overlap
computation and synchronization, the matrix is organized as bands to
make the granularity of the computation adjustable, as demonstrated in
Figure~\ref{fig:Viewbands}.  A task is associated to a band and is
defined as the computation to partially factor the band to the current
frontier.

For each band, the program must remember up to what column this band
has been partially factored. We call this column the {\em current
  position}, which is the start point of factorization for the next
task attached to this band. In addition, it is important to use a
variable to remember the first band that has not been factored
completely.  After the first unfinished band is completely factored,
the frontier global value is increased by the number of rows in the
band. This completely factored band should be broadcast to all
machines in the distributed-memory case or shared by all threads in
the multi-core case.

The smaller the band size, the larger the number of synchronization
points. However, TPILU(k) prefers a smaller band size, that leads to
more parallel tasks.  Moreover, the lower bound of the factorization
time is the time to factor the last band, which should not be very
large. Luckily, shared memory allows for a smaller band size because
the synchronization here is to read/write the frontier, which has a
small cost.

While the strategy of bands is well known to be efficient for dense
matrices (e.g., see~\cite{Cooperman98}), researchers hesitate to use
this strategy for sparse matrices because they may find only a small
number of relatively dense bands, while all other bands are close to
trivial.  The TPILU(k) algorithm works well on sparse matrices because
successive factoring of bands produces many somewhat dense bands (with
more fill-ins) near the end of the matrix.  TPILU(k) uses static load
balancing whereby each worker is assigned a fixed group of bands
chosen round robin so that each thread will also be responsible for
some of the denser bands.

\subsection{Communication in the Distributed Memory Case}
\label{PIPIELINE}
Only the completely factored bands are useful for the factorization of
other bands.  The intermediate result is not truly needed by other
``workers''. This observation, with the static load balancing, helps
to decrease the communication overhead to a minimum by sending an
update message only for each completely factored band.

The static load balancing also produces a more regular communication
that fits well with the pipelining communication of the next
section. A further virtue of this strategy is that it uses a fixed
number of message buffers and a fixed buffer size. This avoids
dynamically allocating the memory for message handling. Under the
strategy of static load balancing, the computations on all processors
are coordinated so as to guarantee that no processor can send two
update messages simultaneously. In other words, a processor must
finish broadcasting an update message before it factors another band
completely.

\paragraph{Pipeline Communication for Efficient Local Network Usage. }
Although the bands can be factored simultaneously, their completion
follows a strict top-down order. When one band is completely factored,
it is best for the node that holds the first unfinished band to
receive the result first. This strategy is implemented by a
``pipeline'' model.  Following this model, all nodes are organized to
form a directed ring.  The message is transferred along the directed
edge. Every node sends the message to its unique successor until each
node has received a copy of the message. After this message is
forwarded, each node uses this message to update its memory data.

\begin{figure}
\begin{center}
\includegraphics[scale=1]{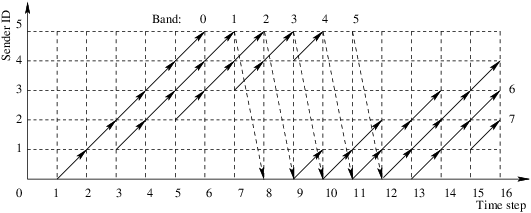}
\caption{``Pipeline'' model.  The horizontal axis is the time
  step. The vertical axis is the sender id.  The lines represent when
  the algorithm sends a message.  The time step and the sender id of
  the source are indicated.  The receiver is always the successor of
  the source.  The message is marked by the corresponding band number.
  Only the first several messages are shown.}
\label{figspread-model}
\end{center}
\end{figure}

Figure~\ref{figspread-model} shows that this model achieves the
aggregate bandwidth. In this figure, the horizontal axis is the time
step while the vertical axis is the sender ID~(the rank number of each
node).  Note that at most time steps, there are several nodes
participating. Each of them is either sending a message to its
successor or receiving a message from its predecessor.

Algorithm~\ref{alg:PILUK} describes the implementation to overlap the
computation with the communication based on the ``pipeline''
model. Both symbolic factorization and numeric factorization can use
this model, except for the case of $k=1$, whose symbolic factorization
reduces to trivial parallelism as discussed in
Section~\ref{sec:EndBaseAlgo}.

\begin{algorithm}
\begin{algorithmic}[1]
\STATE receive from predecessor //non-blocking receive
\STATE //Loop until all bands are factored completely
\WHILE{ firstUnfactoredBand $<$ numberOfBands }
  \STATE get new task (band $ID$) from the ``master'' to work on
  \IF{ there was a band to work on}
    \STATE doTask(band) // factor band using all previous bands
    \IF{ band is not factored completely }
      \STATE //not factored completely, then non-blocking test
      \STATE try to receive a message for some band
      \IF{ a newly factored band is received}
        \STATE send band to successor //non-blocking send
        \STATE update our copy of newly factored band
        \STATE continue to receive and update until our band
               is completely factored
      \ENDIF
    \ELSE
      \STATE send our factored band to successor //non-blocking send
    \ENDIF
  \ELSE
    \STATE wait until a new band is available, while in background continuing
           to receive other factored bands from predecessor, updating our copy,
           and sending the factored band to our successor
  \ENDIF
\ENDWHILE
\end{algorithmic}
\caption{Parallel ILU(k) algorithm with the ``pipeline'' model}
\label{alg:PILUK}
\end{algorithm}

\subsection{Optimized Symbolic Factorization}
\label{sec:EndBaseAlgo}
\noindent {\bf Static Load Balancing and TPMalloc}. Simultaneous
memory allocation for fill-ins is a performance bottleneck for
shared-memory parallel computing. TPILU(k) takes advantage of a
thread-private malloc library to solve this issue as discussed
in~\cite{xingenejohn}.  TPMalloc is a non-standard extension to a
standard allocator implementation, which associates a thread-private
memory allocation arena to each thread.  A thread-local global
variable is also provided, so that the modified behavior can be turned
on or off on a per-thread basis.  By default, threads use
thread-private memory allocation arenas.  The static load balancing
strategy guarantees that if a thread allocates memory, then the same
thread will free it, which is consistent with the use of a
thread-private allocation arena.

\noindent {\bf Optimization for the Case~$k=1$}. When $k=1$, it is
possible to symbolically factor the bands and the rows within each band in
any desired order.  This is because if either $f_{ih}$ or $f_{hj}$ is
an entry of level~$1$, the resulting fill-in~$f_{ij}$ must be an
element of level~$2$ or level~$3$. So $f_{ij}$ is not inserted into
the filled matrix~$F$.  As a first observation, the symbolic
factorization now becomes pleasingly parallel since the
processing of each band is independent of that of any other.

Second, since the order can be arbitrary, even the purely sequential
processing within one band by a single thread can be made more
efficient.  Processing rows in reverse order from last to first is the
most efficient, while the more natural first-to-last order is the
least efficient.  First-to-last is inefficient, because we add
level~$1$ fill-ins to the sparse representation of earlier rows, and
we must then skip over those earlier level~$1$ fill-ins in determining
level~$1$ fill-ins of later rows.  Processing from last to first
avoids this inefficiency.

\subsection{Other Optimizations}
\label{OMCM} 
\noindent
{\bf Optimization on Clusters with Multi-core Nodes.} A hybrid memory model using
multiple computers, each with multiple cores, helps to further improve
the performance for the TPILU(k) algorithm. On each node, start
several threads as ``workers'' and one particular thread as a
``communicator'' to handles messages between nodes. This design leads
to better performance than having each thread worker communicate
directly with remote threads. The reason is that the
``MPI\_THREAD\_MULTIPLE'' option of MPI can degrade performance.

\smallskip
\noindent {\bf Optimization for Efficient Matrix Storage.} The
compressed sparse row format~(CSR) is used in the iteration phase due
to its efficiency for arithmetic operations.  However, the CSR format
does not support enlarging the memory space for several rows
simultaneously.  Therefore, TPILU(k) initializes the matrix $F$ in
row-major format during the symbolic factorization phase.  After the
matrix non-zero pattern is determined by symbolic factorization,
TPILU(k) changes the output matrix $F$ from row-major format back to
CSR format. The format transformations happened during the
factorization phase with a negligible cost.

\subsection{Optional Level-Based Incomplete Inverse Method}
\label{sec:IncompleteInverse}
The goal of this section is to describe the level-based incomplete
inverse method for solving $\widetilde{L}x=p$ by matrix-vector
multiplication: $x=\widetilde{\widetilde{L}^{-1}}p$.  This avoids the
sequential bottleneck of using forward substitution on
$\widetilde{L}x=p$.  We produce incomplete inverses
$\widetilde{\widetilde{L}^{-1}}$ and $\widetilde{\widetilde{U}^{-1}}$
so that the triangular solve stage of the linear solver (i.e., solving
for $y$ in $\widetilde L \widetilde U y=p$ as described in
Equation~\eqref{eq:triangleSolve} of Section~\ref{sec:ReviewILUK}) can
be trivially parallelized
($y=\widetilde{\widetilde{U}^{-1}}\widetilde{\widetilde{L}^{-1}}p$)
while also enforcing bit compatibility.  Although details are omitted
here, the same ideas are then used in a second stage: using the
solution~$x$ to solve for $y$ in $\widetilde U y=x$.

Below, denote the matrix $(-\beta_{it})_{t \le i}$ to be
the lower triangular matrix $\widetilde{L}^{-1}$. Recall that
$\beta_{ii}=1$, just as for $\widetilde L$.  First, we have
Equation~\eqref{forwardsubB}, i.e., $x=\widetilde{L}^{-1}p$.  Second,
we have Equation~\eqref{forwardsubA}, i.e., the equation for solving
$\widetilde{L}x=p$ by forward substitution. Obviously,
Equation~\eqref{forwardsubB} and Equation~\eqref{forwardsubA} define
the same $x$.
\begin{subequations}
\label{forwardsub}
\begin{minipage}{0.5\linewidth}
\begin{equation}
\label{forwardsubB}
 x_i = \sum_{t < i} (-\beta_{it}) p_t + p_i
\end{equation}
\end{minipage}
\begin{minipage}{0.5\linewidth}
\begin{equation}
\label{forwardsubA}
 x_i = p_i-\sum_{h < i} f_{ih}x_h 
\end{equation}
\end{minipage}
\end{subequations}
Substituting Equation~\eqref{forwardsubB} into
Equation~\eqref{forwardsubA}, one has Equation~\eqref{combined}.
\begin{equation}\label{combined}
\begin{split}
x_i &=p_i-\sum_{h < i} f_{ih} \left(\sum_{t < h} (-\beta_{ht}) p_t
    + p_h\right)=\sum_{t<i} \left(-\left(f_{it}
    - \sum_{ t < h < i} f_{ih} \beta_{ht}\right)\,\right)p_t + p_i\\
\end{split}
\end{equation}
Combining the right hand sides of equations~\eqref{forwardsubB}
and~\eqref{combined} yields Equation~\eqref{inversedefine}, the
defining equation for~$\beta_{it}$.
\begin{equation}\label{inversedefine}
\begin{split}
\beta_{it} & =f_{it} - \sum_{ t < h < i} f_{ih} \beta_{ht}\\
\end{split}
\end{equation}
Equation~\eqref{inversedefine} is the basis for computing
$\widetilde{L}^{-1}$ (\hbox{a.k.a. $(-\beta_{it})_{t \le i}$}).
Recall that $f_{ij}$ was initialized to the matrix~$A$.  In algorithm
steps~\eqref{lufactor2} and~\eqref{lufactor} below, row~$i$ is
factored using ILU(k) factorization, which computes $\widetilde L$ and
$\widetilde U$ as part of a single matrix.  These steps are
reminiscent of Gaussian elimination using pivoting
element~$f_{hh}$. Steps~\eqref{lufactor2} and~\eqref{lufactor} are used in
steps~\eqref{inversefactor} and~\eqref{flipfactor} to
compute~$\widetilde L^{-1}$.
\begin{subequations}
\begin{minipage}{0.5\linewidth}
\begin{equation}
\label{lufactor2}
f_{ih} \leftarrow f_{ih}f^{-1} _{hh}
\end{equation}
\end{minipage}
\begin{minipage}{0.5\linewidth}
\begin{equation}
\label{lufactor}
\forall j>h, f_{ij} \leftarrow f_{ij} - f_{ih} f_{hj}
\end{equation}
\end{minipage}
\begin{minipage}{0.5\linewidth}
\begin{equation}
\label{inversefactor}
 \forall t<h, f_{it} \leftarrow f_{it} - f_{ih}f_{ht}
\end{equation}
\end{minipage}
\begin{minipage}{0.5\linewidth}
\begin{equation}
\label{flipfactor}
 \forall t<i, f_{it} \leftarrow -f_{it}
\end{equation}
\end{minipage}
\end{subequations}

The matrix $\widetilde{L}^{-1}$ is in danger of becoming dense. To
maintain the sparsity, we compute the level-based incomplete inverse
matrix $\widetilde{\widetilde{L}^{-1}}$ following the same non-zero
pattern as $\widetilde{L}^{-1}$. The computation for
$\widetilde{\widetilde{L}^{-1}}$ can be combined with the original
numeric factorization phase.  A further factorization phase is added to
compute~$\widetilde{\widetilde{U}^{-1}}$ by computing matrix entries
in reverse order from last row to first and from right to left within
a given row.

Given the above algorithm for $\widetilde{\widetilde{L}^{-1}}$ and a
similar algorithm for $\widetilde{\widetilde{U}^{-1}}$, the triangular
solve stage is reduced to matrix-vector multiplication, which can be
trivially parallelized.  Inner product operations are not parallelized
for two reasons: first, even when sequential, they are fast; second,
parallelization of inner products would violate bit-compatibility by
changing the order of operations.

\section{Experimental Results}
\label{sec:Experiment}
We evaluate the performance of the bit-compatible parallel
ILU(k) algorithm, TPILU(k), by comparing with two commonly used parallel
preconditioners, PILU~\cite{Hysom00ascalable} and
BJILU~\cite{SAADBOOK} (Block Jacobi ILU(k)).
Both PILU and BJILU are based on {\em domain
  decomposition}.  Under the framework of
Euclid~\cite[Section~6.12]{HYPREUSERMANUAL}, both preconditioners
appear in Hypre~\cite{HYPREUSERMANUAL}, a popular linear solver package
under development at Lawrence Livermore National Laboratory since 2001.

The primary test platform is a computer with four Intel Xeon E5520
quad-core CPUs (16~cores total).  Figure~\ref{fig:TPILUcompare}
demonstrates the scalability of TPILU(k) both on this primary platform
and a cluster including two nodes connected by Infiniband.  Each node has a
single Quad-Core AMD Opteron 2378~CPU.  The operating system is
CentOS~5.3 (Linux~2.6.18) and the compiler is gcc-4.1.2 with the
``-O2'' option. The MPI library is OpenMPI~1.4.  Within Hypre,
the same choice of iterative solver is used to test both Euclid
(PILU and BJILU) and TPILU(k).  The chosen iterative solver is
preconditioned stabilized bi-conjugate gradients with the default
tolerance $rtol=10^{-8}$.  
Note that the Euclid framework employs multiple MPI processes
communicating via MPI's shared-memory architecture, instead of
directly implementing a single multi-threaded process.

\subsection{Driven Cavity Problem}
\begin{figure}
\centering
\subfloat[Euclid for $e20r3000$]
{\label{fig:olde20r3000}\includegraphics[scale=1]{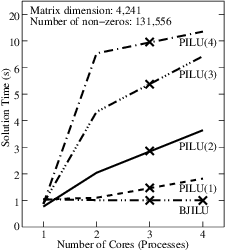}}
~
\subfloat[Euclid for $e30r3000$]
{\label{fig:olde30r3000}\includegraphics[scale=1]{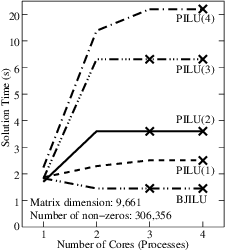}}
~
\subfloat[Euclid for $e40r3000$]
{\label{fig:olde40r3000}\includegraphics[scale=1]{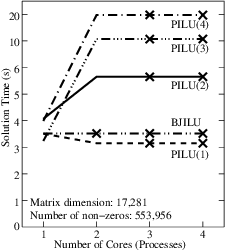}}
\caption{Euclid PILU and BJILU for Driven Cavity Problem using a
  Single AMD Opteron ($4$ Cores). ``X'' means fail, and the time is
  arbitrarily shown to be an interpolated value or the same as for the
  preceding number of threads.  Note that in
  Figure~\ref{EUCLIDDCP}(a), PILU(k) actually breaks down for
  3~threads, while then succeeding for 4~threads.}
\label{EUCLIDDCP}
\end{figure}

This set of test cases~\cite{RandomMatrix} consists of some difficult
problems from the modeling of the incompressible Navier-Stokes
equations. These test cases are considered here for the sake of
comparability.  They had previously been chosen to demonstrate the
features of PILU by~\cite{HysomPothen03}. Here, we test on three
representatives: $e20r3000$, $e30r3000$ and $e40r3000$.
Figure~\ref{EUCLIDDCP} shows that both Euclid PILU and Euclid BJILU
are influenced by the number of processes and the level~$k$ when
solving driven cavity problems. With more processes or larger~$k$,
both the PILU and BJILU preconditioners tend to slow down, break down
or diverge.

\begin{figure}
\centering
\subfloat[Matrix $e20r3000$]
{\label{fig:newe20r3000}\includegraphics[scale=1]{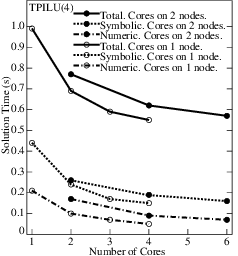}}
~
\subfloat[Matrix $e30r3000$]
{\label{fig:newe30r3000}\includegraphics[scale=1]{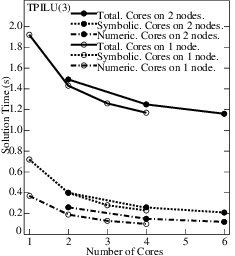}}
~
\subfloat[Matrix $e40r3000$]
{\label{fig:newe40r3000}\includegraphics[scale=1]{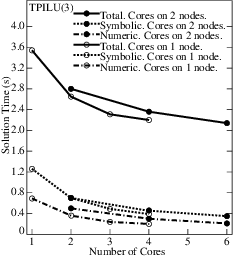}}
\caption{TPILU(k) for the Driven Cavity Problem Using 2 AMD Opteron
  ($2 \times 4$ Cores).  The experimental runs for 1,2,3,4~threads are
  all for a 4-core shared memory CPU.  The experimental runs for
  2,4,6~threads are all for two nodes with 4-cores per node, while an
  additional thread per node is reserved for communication between
  nodes in order to replicate bands. }
\label{fig:TPILUcompare}
\end{figure}

Euclid registers its best solution time for $e20r3000$ by using
PILU(2) with 1~process, for $e30r3000$ by using BJILU with
2~processes, and for $e40r3000$ by using PILU(1) with 2~processes.
The reason that Euclid PILU obtains only a small speedup for these
problems is that PILU requires the matrix to be {\em
  well-partitionable}, which is violated when using a larger level~$k$
or when employing more processes. Similarly, Euclid BJILU must
approximate the original matrix by a number of subdomains equal to the
number of processes. Therefore, higher parallelism forces BJILU to
ignore even more off-diagonal matrix entries with more blocks of
smaller block dimension, and eventually the BJILU computation just
breaks down.

In contrast, TPILU(k) is bit-compatible. Greater parallelization only
accelerates the computation, while also never introducing
instabilities or other negative side effects.
Figure~\ref{fig:newe20r3000} illustrates that for the $e20r3000$ case,
TPILU with level $k=4$ and 4~threads leads to a better performance
(0.55~s) than Euclid's 0.78~s (Figure~\ref{fig:olde20r3000}).  For the
$e30r3000$ case, TPILU(k) finishes in~1.16~s
(Figure~\ref{fig:newe30r3000}), as compared to 1.47~s for BJILU and
1.64~s for PILU (Figure~\ref{fig:olde30r3000}).  For the $e40r3000$
case, TPILU(k) with $k=3$ finishes in~2.14~s
(Figure~\ref{fig:newe40r3000}), as compared to~3.15~s for PILU and
3.52~s for BJILU
(Figure~\ref{fig:olde40r3000}). Figure~\ref{fig:newe40r3000}
demonstrates the potential of TPILU(k) for further performance
improvements when a hybrid architecture is used to provide additional
cores: the hybrid architecture with 6~CPU cores over two nodes
connected by Infiniband is even better~(2.14~s) than the shared-memory
model with a single quad-core CPU~(2.20~s).

\subsection{3D 27-point Central Differencing}
\begin{figure}
\centering
\subfloat[Euclid PILU]
{~~\label{fig:hypre404040}\includegraphics[scale=1]{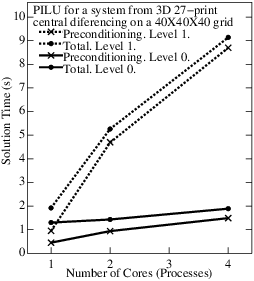}~~}
\subfloat[Comparison of Euclid PILU and TPIILU]
{~~~~~~~~~\label{fig:new3D}\includegraphics[scale=1]{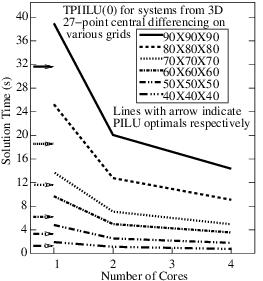}~~~~~~~~~}
\caption{Solving Linear System from 3D 27-point Central Differencing on
  Grid using a Single AMD Quad-Core Opteron.  Focusing on the algorithm only, the comparison ignores reusing the domain decomposition over multiple linear system solutions.
}
\label{fig:3D27point}
\end{figure}

As pointed out in~\cite{HysomPothen03}, ILU(k) preconditioning is
amenable to performance analysis since the non-zero patterns of the
resulting ILU(k) preconditioned matrices are identical for any partial
differential equation (PDE) that has been discretized on a grid with a
given stencil. However, a parallelization based on domain
decomposition may eradicate this feature since it generally relies on
re-ordering to maximize the independence among subdomains.  The
re-ordering is required for domain decomposition since it would
otherwise face a higher cost dominated by the resulting denser
matrix. As Figure~\ref{fig:hypre404040} shows, Euclid PILU degrades
with more processes when solving a linear system generated by
3D~27-point central differencing for Poisson's equation. The
performance degradation also increases rapidly as the level~$k$ grows.

This performance degradation is not an accident.
The domain-decomposition computation
dominates when the number of non-zeros per row is
larger (about 27 in this case). Therefore, the sequential algorithm with
the level $k=0$ wins over the parallelized PILU in the contest for the best
solution time. This observation holds true for all grid sizes tested:
from $50 \times 50 \times 50$ to $90 \times 90 \times 90$.  In
contrast, for all of
these test cases, TPIILU (the level-based incomplete inverse submethod
of TPILU(k)) leads to improved performance using 4~cores, as seen in
Figure~\ref{fig:new3D}.

\subsection{Model for DNA Electrophoresis: cage15}
\begin{figure}
\centering
\subfloat[TPIILU(0) for cage15]
{\label{fig:cage15}\includegraphics[scale=1]{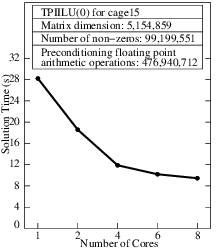}}
~~~~
\subfloat[TPIILU(1) for ns3Da]
{\label{fig:ns3Dathread}\includegraphics[scale=1]{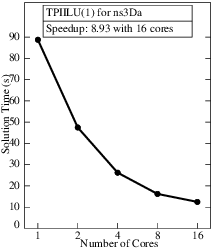}}
~~~~
\subfloat[TPMalloc Performance]
{\label{fig:tpmalloc}\includegraphics[scale=1]{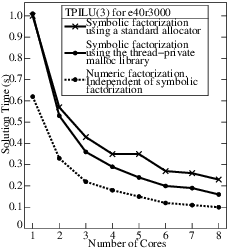}}
\caption{TPIILU(k)/TPILU(k) using 4 Intel Xeon E5520 ($4 \times 4$ Cores)}
\label{fig:otherApps}
\end{figure}

The cage model of DNA electrophoresis~\cite{RandomMatrix23} describes
the drift, induced by a constant electric field, of homogeneously
charged polymers through a gel. We test on the largest case in this
problem set:~$cage15$. For $cage15$, TPIILU(0) obtains a speedup
of~2.93 using 8~threads~(Figure~\ref{fig:cage15}).  The ratio of the
number of FLoating point arithmetic OPerations (FLOPs) to the number
of non-zero entries is less than~5. This implies that ILU(k)
preconditioning just passes through matrices with few FLOPs. In other
words, the computation is too ``easy'' to be further sped up.

\subsection{Computational Fluid Dynamics Problem: ns3Da}
The problem $ns3Da$~\cite{RandomMatrix23} is used as a test case in
FEMLAB, developed by Comsol, Inc. Because there are zero diagonal
elements in the matrix, we use TPIILU with level~$k=1$ as the
preconditioner. Figure~\ref{fig:ns3Dathread} shows a speedup of~8.93
with 16~threads since the preconditioning is floating-point intensive.

\subsection{TPMalloc Performance}
For a large level~$k$, the symbolic factorization time will
dominate. To squeeze greater performance from this first phase,
glibc's standard malloc is replaced with a thread-private malloc
(TPMalloc). Figure~\ref{fig:tpmalloc} demonstrates that the
improvement provided by TPMalloc is significant whenever the number of
cores is greater than~2.

\subsection{Results to Distributed Memory}

\subsubsection{ns3Da for distributed memory.} 
We also test the ns3Da case on a cluster consisting of 33~nodes. Each
node of this cluster has 4~CPU cores (dual processor, dual-core),
2.0~GHz Intel Xeon EM64T processors with either 8~GB or 16~GB per
node.  The nodes are connected by a Gigabit Ethernet network. Each
node is configured with Linux~2.6.9, gcc~3.4.5 and MPICH2~1.0 as the
MPI library. On this cluster, TPILU(k) obtains a speedup of 7.22~times
(Table~\ref{tbl:ns3DaMPI}) for distributed memory and 7.82~times
(Table~\ref{tbl:ns3DaMPIMT}) for hybrid memory.

\begin{table}[hbt]
\begin{center}
\begin{tabular}{|c||r|r|r|r|}
\hline
Procs & PC time (s) & \# Iters & Time (s) & Speedup \\
\hline
1&82.26&34&6.39&1\\
2&42.57&34&5.27&1.85\\
4&22.90&34&4.00&3.30\\
8&13.69&34&3.30&5.22\\
16&9.19&34&3.08&7.22\\
\hline
\end{tabular}
\caption{TPILU(k) for $ns3Da$ with 16~nodes on a cluster}
\label{tbl:ns3DaMPI}
\end{center}
\end{table}

\begin{table}[hbt]
\begin{center}
\begin{tabular}{|c||r|r|r|r|r|}
\hline
Procs & Threads & PC time (s) & \# Iters & Time (s) & Speedup \\
\hline
1&1&82.26&34&6.39&1\\
2&6&15.28&34&2.83&4.90\\
4&12&11.39&34&2.63&6.32\\
8&24&8.59&34&2.75&7.82\\
%20&8.40s&34&3.08s&\\
\hline
\end{tabular}
\caption{TPILU(k) for $ns3Da$ with 8~nodes on a cluster with
  1~dedicated communication thread and 3~worker threads per
  process and one process per node}
\label{tbl:ns3DaMPIMT}
\end{center}
\end{table}

\subsubsection{Further distributed memory experiments based on matgen. }
This section highlights TPILU(k) on distributed memory for the
factorization of sparse matrices generated by matgen~\cite{Matgen}. A
copy of the input matrix $A$ is assumed to reside on each node of the
cluster before the computation.  After the computation, the result
matrix $F$ also resides on each node of the
cluster. See~\cite{XINDONG0803} for a more detailed discussion
concerning TPILU(k) on clusters.

\subsubsection{Symbolic factorization versus numeric factorization. }
\begin{figure}
\begin{center}
\includegraphics[scale=1.4]{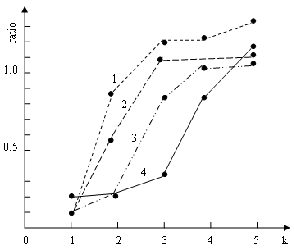}
\caption{Comparison of symbolic factorization and numeric factorization (sequential algorithm); LEVEL $k = 1, 2, 3, 4$ and $5$; 1 is the result for a matrix of $1K \times 1K$; 2 is the result for a matrix of $2K \times 2K$; 3 is the result for a matrix of $4K \times 4K$; 4 is the result for a matrix of $8K \times 8K$;}
\label{sequentiallufig}
\end{center}
\end{figure}

This set of experiments illustrate how the ratio of symbolic
factorization to numeric factorization changes when $k$ increases
gradually.  Figure~\ref{sequentiallufig} demonstrates the computation
time of four matrices measured for both symbolic factorization and
numeric factorization. The initial matrix densities are $0.073$,
$0.036$, $0.009$ and $0.002$ respectively. In all cases, the ratio of
symbolic factorization to numeric factorization does not decrease when
the level $k$ grows from~$1$ to~$5$. When $k$ is large enough, the
ratio goes beyond 1.

Therefore, the time for symbolic factorization is almost the same as
or even a little greater than that of numeric factorization if no
entry is skipped in the first phase (turning off the optimization as
described in Section~\ref{OMCM}). Although the non-zero elements are
inserted dynamically and this results in fewer comparisons, the
insertion of an entry is costly. To insert an entry, we move the
remaining entries in order to open enough space for the new fill
entry.

{\em Remark: The symbolic factorization for the case $k=1$ is
  lightweight due to the optimization mentioned in Section~\ref{OMCM}.
  Consider a matrix of dimension 20,000 and a level limit of~1 in one
  of our experiments.  After symbolic factorization, the number of
  entries is 1,239,058 and all of them are involved in numeric
  factorization. On the other hand, only 265,563 level~0 elements
  cause a new entry in the symbolic factorization phase.}

\subsubsection{Parallel $ILU(k)$ for larger~$k$. }
\label{sec:Experiment3a}
\begin{figure}
\begin{center}
\includegraphics[scale=1.4]{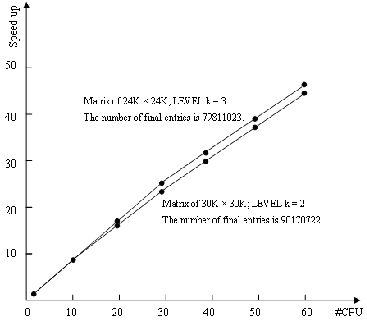}
\caption{Computation for higher level; The matrix densities are 0.00061 and 0.00089.}
\label{fig:parallelddhigherk}
\end{center}
\end{figure}

For the situation $k=2$ and $k=3$, the speedup is good, as
demonstrated by the experimental results in
Figure~\ref{fig:parallelddhigherk}.  For the $24K$~matrix with initial
density 0.00061, $k=3$ creates 79,811,023 final entries. For the $30K$
matrix with initial density 0.00089, $k=2$ results in 90,170,722 final
entries. For both situations, we achieve nearly linear speedup through
the largest example: 60~CPU cores. This result is reasonable, since
raising~$k$ to~2 or 3 causes more fill-ins, and a denser result
matrix.  The added floating-point arithmetic per task leads to a
relatively larger computation sufficient to overlap communication.

\subsubsection{Case $k=1$ : $PILU(1)$. }
\label{sec:Experiment3b} This is the most commonly used case. The
results of the sequential ILU(k) are collected in
Table~\ref{tbl:sequentialdd} while TPILU(1) is presented in
Table~\ref{tbl:paralleldd}. For $k = 1$, all causative entries are
initial entries with level~0. So in the first pass, all computation
that enlarges the size of filled matrix or updates the level is
involved in merely initial entries. There is no communication in the
first pass. The final number of entries determines the communication
overhead of a single machine for the second pass. The final entry
number, density and non-zero pattern determine the amount of
computation in the second pass.

\begin{table}[hbt]
\begin{center}
\begin{tabular}{|c||c|c|c|c|}
\hline
n & \#Initial entry & \#Final entry & Time (s)\\
\hline
\hline
40K & 5120950 & 196223519 &  445.2 + 8938.2\\
\hline
80K & 6960983 & 195202037 &  234.0 + 4120.8\\
\hline
160K & 9832794 & 198969083 & 140.2 + 2112.7 \\
\hline
320K & 14090553 & 206489590 & 93.4 + 1162.0 \\
\hline
\end{tabular}
\caption{Computation of sequential algorithm; Level $k=1$. {\bf (The matrix densities are 0.003, 0.001, 0.00037 and 0.00013.)}
}
\label{tbl:sequentialdd}
\end{center}
\end{table}

\begin{table}[hbt]
\begin{center}
\begin{tabular}{|c||c|c|c|c|}
\hline
n & \#CPU Core & \#Band & Time (s) & Speedup \\
\hline
\hline
40K & 50 & 20480 & 9.5 + 217.4 & 41.4 \\
\hline
40K & 60 & 20480 & 7.8 + 191.6 & 47.1 \\
\hline
80K & 40 & 20480 & 6.7 + 142.9 & 29.1 \\
\hline
80K & 60 & 20480 & 5.0 + 98.7 & 42.0 \\
\hline
160K & 30 & 40960 & 4.7 + 101.8 & 21.2 \\
\hline
160K & 60 & 40960 & 2.5 + 64.4 & 33.7 \\
\hline
320K & 30 & 81920 & 3.3 + 89.8 & 13.5 \\
\hline
320K & 40 & 81920 & 2.4 + 71.3 & 17.0 \\
\hline
320K & 60 & 81920 & 1.6 + 59.0 & 20.7 \\
\hline
\end{tabular}
\caption{Computation for Level $k=1$. {\bf (The matrix densities are 0.003, 0.001, 0.00037 and 0.00013.)}
}
\label{tbl:paralleldd}
\end{center}
\end{table}

From Tables~\ref{tbl:sequentialdd} and~\ref{tbl:paralleldd}, one can see
the following: for a matrix of dimension 40,000 with density~0.003 and
a matrix of dimension 80,000 with density~0.001, TPILU(k) obtains a
sub-linear speedup for 60~CPU cores.  For a matrix of dimension
160,000 with density~0.00037, and a matrix of dimension 320,000 with
density 0.00013, TPILU(k) achieves a maximum speedup at 60~CPU cores.

The symbolic factorization phase always obtains a linear speedup
because there is no communication overhead. However, the numeric
factorization phase does not achieve a linear speedup for all cases.
The decreasing speedup in the numeric factorization phase is the major
part that influences the total speedup. It is explained as follows.

Suppose the matrix has $n_f$~entries finally.  Following the
``pipeline'' model (Section~\ref{PIPIELINE}), both the column number
and value of each entry in each band are sent to a unique child node
once except for the bands that are handled by the child
node. Therefore, the communication overhead is about $8n_fB$ per node.

Considering the four matrices in the above experiments, the number of
final entries is $200M$ in all cases. So the communication overhead is
about $200M \times 8B$ per node in the second phase. As the matrix
dimension increases, the density decreases, as does the amount of
floating-point arithmetic operations and the computation-communication
ratio. Also the optimal number of CPU cores then decreases.

Increasing the number of CPU cores decreases the
computation-communication ratio by increasing the total communication
overhead and decreasing the computation burden of each machine. In
order to improve the speedup further, it is necessary to decrease the
total communication time to overlap communication and computation
well. One solution is to increase bandwidth using a high performance
cluster lonestar.tacc.utexas.edu, which has more bandwidth.

\begin{figure}
\begin{center}
\includegraphics[scale=1.4]{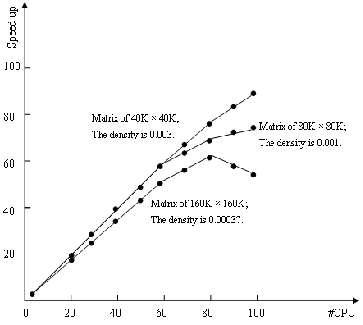}
\caption{Computation on Lonestar; Level $k=1$}
\label{fig:paralleldd2}
\end{center}
\end{figure}

The Lonestar cluster is configured with 5200 compute-node processors
connected by a 10-Gigabyte network. Each node has two Xeon Intel
Duo-Core 64-bit processors (totally 4 cores) and 8 GB memory. The core
frequency is 2.66GHz. The operating system is Linux 2.6 and the
compiler is Intel 9.1. The MPI library is
MVAPICH. Figure~\ref{fig:paralleldd2} presents experimental results on
Lonestar using the same matrices of 40K, 80K and 160K as in the
previous experiment. It demonstrates the scalability of TPILU(k)
toward 80--100 CPU cores given sufficient bandwidth.

\subsubsection{Scalability to Grid. }
\label{sec:Experiment4}
TPILU(k) also has reasonable latency tolerance. This is important for
such architectures as the Computational Grid. TPILU(k) manages to
overcome the large latency of inter-cluster, which is generally a few
$ms$ compared to a few $\mu s$ inter-cluster latency.  To test
TPILU(k) on the Grid, we simulate the communication latency of the
Grid on a cluster by adding some delay ($\ge 17.5ms$, this number
being chosen because our tests of the round-trip time on the Internet
are always less than $35ms$) before message sending for each gateway
node.

A $32K \times 32K$ matrix with the initial density 0.00458 is
generated for this experiment. The number of final entries is
210,212,433 after symbolic factorization.  The results are collected
in Table~\ref{tbl:movetogrid}. In this table, the number of CPU cores
is expressed as the number of clusters times the number of CPU cores
of each cluster. Table~\ref{tbl:movetogrid} demonstrates a single
cluster of 100~nodes exhibiting 64 times speedup. Two clusters of
50~nodes each caused the speedup to degrade to only 45 times (for a
typical 17~ms delay over the Internet), and to a 42~times speedup for
a 24~ms long delay.  Adding a third cluster of 50~nodes contributed
almost no additional speedup. Table~\ref{tbl:movetogrid} illustrates
the following features of TPILU(k) on Grid:
\begin{enumerate}
\item The average bandwidth of communication on the Grid is not as good as
  that on cluster due to the communication latency. This always makes
  the speedup decrease more or less. The maximal speed up is at 100
  CPU cores without latency.

\item The speedup decreases if the latency increases given the same
  amount of clusters and CPU cores. The algorithm performs better with
  $17ms$ latency than with $24ms$ latency assuming other parameters
  are same.

\item The performance of TPILU(k) does not deteriorate. In both $17ms$
  and $24ms$ cases, the maximal speedup is obtained by the most CPU cores.
\end{enumerate}

The reason that TPILU(k) maintains good performance on the Grid is
that this algorithm supports a strategy to mitigate the influence of
latency via using as small a number of bands as possible given there
are still enough number of bands for load-balancing. The fewer-bands
policy reduces the number of messages required for band replication.

\begin{table}[hbt]
\begin{center}
\begin{tabular}{|c||c|c|c|c|}
\hline
 Delay(ms)& \#CPU Core & \#Band & Time (s) & Speedup \\
\hline
\hline
 0 & 1 & & 7821.3 & \\
\hline
 0 & 60 & 4096 & 7.9 + 203.2 & 37.1 \\
\hline
 0 & 100 & 8192 & 7.1 + 115.1 & 64 \\
\hline
 17 & $2 \times 50$ & 2048 & 5.3 + 169.2 & 44.8 \\
\hline
 24 & $2 \times 50$ & 2048 & 5.0 + 182.8 & 41.6 \\
\hline
 17 & $2 \times 60$ & 2048 & 4.8 + 158.0 & 48.0 \\
\hline
 24 & $2 \times 60$ & 2048 & 4.8 + 169.4 & 44.9 \\
\hline
 17 & $3 \times 50$ & 1024 & 3.8 + 159.7 & 47.8 \\
\hline
 24 & $3 \times 50$ & 1024 & 3.7 + 163.7 & 46.7 \\
\hline
 17 & $3 \times 60$ & 4096 & 3.1 + 159.6 & 48.1 \\
\hline
\end{tabular}
\caption{Simulation of the performance on Grid; $32K \times 32K$; Level $k=1$. {\bf (The matrix density is 0.00458)}
}
\label{tbl:movetogrid}
\end{center}
\end{table}

\subsection{\bf Experimental Analysis} 
Given a denser matrix, or a higher level $k$ or more CPU cores, the
time for domain-decomposition based parallel preconditioning using
Euclid's PILU(k) can dominate over the time for the iterative solving
phase.  This degrades the overall performance, as seen both in
Figure~\ref{fig:hypre404040} and in Figures~\ref{EUCLIDDCP}(a,b,c).  A
second domain-decomposition based parallel preconditioner, Euclid's
BJILU, generally produces a preconditioned matrix of lower quality
than ILU(k) in Figure~\ref{EUCLIDDCP}(a,b,c).  This happens because it
ignores off-diagonal non-zero elements. Therefore, where Euclid
PILU(k) degrades the performance, it is not reasonable to resort to
Euclid BJILU.  Figures~\ref{fig:olde20r3000} and~\ref{fig:olde40r3000}
show that the lower quality of BJILU-based solvers often performed
worse than PILU(k).  Figure~\ref{fig:TPILUcompare} shows TPILU(k) to
perform better than either while maintaining the good scalability
expected of a bit-compatible algorithm.  TPILU(k) is also robust
enough to perform reasonably even in a configuration with two
quad-core nodes.  Additionally, Figures~\ref{fig:new3D}
and~\ref{fig:otherApps} demonstrate very good scalability on a variety
of applications when using the optional level-based incomplete inverse
optimization.

The crucial issue for TPILU(k) on distributed memory is how well this
algorithm is able to overlap computation and communication. For $k=2$
and $k=3$, TPILU(k) achieves nearly linear speedup since computation
dominates communication due to more floating-point arithmetic per
task, as Figure~\ref{fig:parallelddhigherk} shows.  For $k=1$,
Tables~\ref{tbl:sequentialdd} and~\ref{tbl:paralleldd} illustrate that
the optimized version of TPILU(1) produces a 21-fold speedup on a
departmental cluster (Gigabit Ethernet) over 30 nodes, operating on a
matrix of dimension 160,000 and density~0.00037. With a high
performance cluster (InfiniBand interconnect), TPILU(k) registers a
58-fold speedup with 60 nodes operating on a matrix of dimension
80,000 and density~0.001, as Figure~\ref{fig:paralleldd2}
demonstrates. Table~\ref{tbl:movetogrid} highlights the potential of
TPILU(k) on the Grid even with a large communication delay for
inter-cluster message passing: although the TPILU(k) speedup degrades,
we still observe an overall performance improvement with two and three
clusters participating in a computation.

\section{Related Work}
\label{RELATEDWORK}
There are many sequential preconditioners based on incomplete LU
factorization.  Two typical sequential preconditioners are
ILUT\cite{SAADBOOK} and
ILU(k)\cite{springerlink:10.1007/BF01931691}. ILU(k) was formalized to
solve the system of linear equations arising from finite difference
discretizations in 1978.  In 1981, ILU(k) was extended to apply to
more general problems~\cite{WATTS}.  A recent sequential
preconditioner is ILUC~\cite{ILUC2,ILUC3}.

In~\cite{KarypisKumar97}, a parallel version of ILUT is given for
distributed memory parallel computers.  However, the parallelism in
this paper comes from the analysis of a special non-zero pattern for a
sparse matrix and does not have high scalability for a general sparse
matrix.

In the process of parallelizing ILU(k) preconditioners, we are faced
with a natural problem: why is it so difficult to speed up ILUT or
ILU(k) when k is small?  We observe that ILU(k) preconditioning is the
kind of computation that accesses lots of memory while using
relatively little floating-point arithmetic in the case of a huge
sparse matrix of lower density with $k=1$ or $k=2$. Therefore, it is
limited by either the memory bandwidth for the shared-memory case or
the network bandwidth for the distributed-memory case when
parallelizing and speeding up an ILU preconditioner with more CPU
cores. Many discussions in~\cite{Benzi2002,DUFF,Michael} contribute
valuable ideas that help us to handle this problem and design a
scalable algorithm.

In~\cite{Fu2,Fu,BIN}, an LU factorization algorithm for distributed
memory machines is implemented. However, this implementation needs a
special API to update and synchronize the distributed memory. It is an
evidence that communication in the distributed memory model is a
bottleneck even for LU factorization when huge sparse matrices are
considered. It implies that the parallelization of ILU(k)
preconditioner is challenging on clusters.  However, it is important
because cluster systems are the mainstream of supercomputers: More
than $70\%$ of all supercomputers in the 2007 TOP500
list~\cite{TOP500} are cluster systems.

In~\cite{XiaoLi96}, the supernode data structure is used to reorganize
a sparse matrix. Those supernodes can be processed in parallel.
Observing that many rows have a similar non-zero pattern in the result
matrix of LU factorization, rows with a similar non-zero pattern can
be organized as one supernode.

The parallel ILU preconditioner~\cite{PBILUM} aims toward distributed
sparse matrices. PMILU~\cite{PMILU} presents a new technique to
reorder a linear system in order to expose greater parallelism. They
represent a category of parallel algorithms based on
reordering. In~\cite{XiaoLi2007}, pivoting is employed to reduce the
number of fill-ins for LU factorization. Similarly, pivoting is used
to improve the robustness of the ILU preconditioner
in~\cite{JZhang2001}. The work in~\cite{duff01algorithms} provides an
algorithm to make the diagonal elements of a sparse matrix large. The
methodology is to compute the optimal pivoting and preprocess a sparse
matrix. If the preprocessing makes the sparse matrix break-down free
for ILU(k) preconditioning, then it is possible to relax the
diagonal dominance condition required by TPILU(k).

Some other previous parallel ILU(k) preconditioners
include~\cite{Hysom00ascalable,ANDERSON88parallelimplementation,Heroux91aparallel}.
The latter two methods, whose parallelism comes from level/backward
scheduling, are stable and were studied in the 1980's and achieved a
speedup of about 4 or~5 on an Alliant FX-8~\cite[1st edition, page
351]{SAADBOOK} and a speedup of 2 or~3 on a Cray~Y-MP.  The more
recent work~\cite{Hysom00ascalable} is directly compared with in the
current work, and is not stable.

\section{Conclusion}
\begin{figure}
\begin{center}
\includegraphics[scale=1]{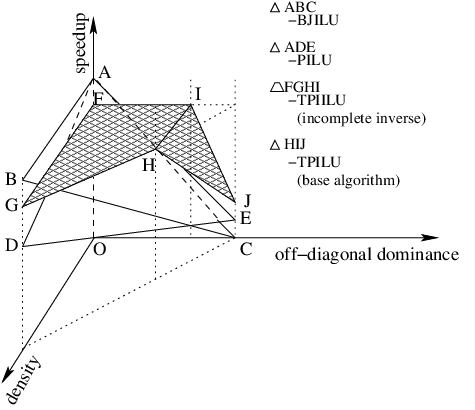}
\caption{Three-dimensional comparison of BJILU, PILU and
  TPILU(k). Higher points represent greater speedup. Off-diagonal
  dominance is intended to express a qualitative rather than
  quantitative concept: how many off-diagonal elements are there that
  are significantly far from zero. TPILU(k)/TPIILU(k) is generally
  faster, but BJILU is faster for matrices with few or small
  off-diagonal elements. PILU(k) is also faster in the above case--but
  only for very sparse matrices.}
\label{FEATURECOMPARISON}
\end{center}
\end{figure}

This work can be graphically summarized via
Figure~\ref{FEATURECOMPARISON}.  While there are regions in which each
of BJILU, PILU and TPILU(k) is faster, TPILU(k) remains competitive
and stable in most cases.  TPILU(k) is more stable than BJILU, as
demonstrated in Figure~\ref{FEATURECOMPARISON}: BJILU has excellent
performance on some types of linear systems, while failing to converge
on other linear systems.  Figure~\ref{FEATURECOMPARISON} also shows
that TPILU(k) is more stable than PILU, whose performance degrades
along two dimensions: density and off-diagonal dominance. 

Even though it is omitted by Figure~\ref{FEATURECOMPARISON}, the
performance degradation increases for both BJILU and PILU when they
employ more threads. In contrast, the bit-compatibility of TPILU(k)
guarantees performance improvement for solving most classes of large
sparse linear systems while maintaining stability.

\section{Acknowledgement}
We acknowledge helpful discussions with Ye Wang at an early stage of
this work.

\bibliography{main}
\bibliographystyle{splncs}
\end{document}